\documentclass[11pt]{article}

\let\ssection=\section
\renewcommand{\section}{\setcounter{equation}{0}\ssection}
\usepackage{bm}
\usepackage{amsmath,dsfont}

\usepackage{graphics,amssymb,amsmath,amsthm,amscd,array}

\def\parag{\hfil\break} 
\def\kikezd{\parag\underbar}

\def\p{{\partial}}
\def\vr{{{\bf r}}}

\def\vk{{{\bf k}}}

\def\vJ{{{\bf J}}}
\def\vl{{{\bf l}}}
\def\vz{{\hat{\bf z}}}
\def\bLambda{{{\bf \Lambda}}}
\def\cD{{{\cal D}}}
\def\vK{{{\bf K}}}

\def\vv{{{\bf v}}}

\def\vA{{{\bf A}}}
\def\vsigma{{{\bm\sigma}}}
\def\vpi{{{\bm\pi}}}

\def\SU{{\rm SU}}
\def\U{{\rm U}}

\def\smallover#1/#2{\hbox{$\textstyle\frac{#1}{#2}$}}
\def\2{{\smallover1/2}}

\def\dAlembert{\vcenter {
    \hbox {\vrule height8pt width0.4pt depth0.0pt
           \vrule height8pt width7.2pt depth-7.6pt
           \vrule height8pt width0.4pt depth0.0pt
           \kern-8pt
           \vrule height0.4pt width8pt depth0.0pt
          \,}}}

\def\rot{{\rm rot\,}}

\title{\bf Non-relativistic scattering of a spin-1/2 particle
off a self-dual monopole
}

\author{L. Feh\'er
\footnote{Bolyai Institute, University of Szeged. Present address:
Research Institute for Particle and Nuclear Physics, Budapest,
Hungary. e-mail: lfeher-at-rmki.kfki.hu}
 \quad and \quad P. A.
Horv\'athy \footnote{Department de Math\'ematiques et d'Informatique, Universit\'e de Metz.
 Present address: Laboratoire de Math\'ematiques et de
Physique Th\'eorique, Universit\'e de Tours, France. e-mail:
horvathy-at-lmpt.univ-tours.fr} }
\begin{document}

\maketitle

\begin{abstract}
The non-relativistic scattering of a spin-1/2 particle off a self-dual monopole reduces, for
large distances, to the dyon problem, studied previously by D'Hoker and Vinet.
The $S$ matrix (calculated by Zwanziger's algebraic method based on the ${\mathrm{o}}(3,1)\oplus{\mathrm{o}}(3)$ dynamical
symmetry, discovered by D'Hoker and Vinet) is shown to factorize into the
product of the spinless $S$-matrix, $S_0$, with a spin-dependent factor. The total cross-section is identical to the one found in the spinless case.
\end{abstract}
\bigskip\noindent
Published as {\sl Mod. Phys. Lett.} {\bf A3} No 15.  1451-1460 (1988).
\bigskip

\goodbreak

The motion of fermions in a monopole field has been extensively
studied \cite{1,2,3,4,5}, in particular in connection with the Rubakov effect \cite{4}.
For a pure Dirac monopole, the radial Hamiltonian is not self-adjoint, and the solution depends
on the boundary condition at $r=0$. For a non-Abelian monopole with symmetry-breaking `Higgs'
potential \cite{6}, the picture is essentially the same: the Higgs field approaches
its asymptotic value as $e^{-\mu r}$ (where $\mu$ is the Higgs mass), and outside the monopole core we are left with a pure gauge monopole, which behaves as an Abelian object \cite{6}.

The situation changes dramatically in the Prasad-Sommerfield limit \cite{6} of vanishing Higgs potential: the long-range massless Higgs field produces an attractive force (also responsible for the existence of static, self-dual,
`Bogomolny' multimonopoles \cite{7}).  The test particles can not reach the
origin, curing the non-self-adjointness of the Hamiltonian. The Dirac equation has bound
state solutions \cite{5}.
In particular, there is an interesting $E=0$ ground state.
For large $r$, the exponential terms can be dropped, the Jackiw-Rebbi equations
can be solved exactly; electric charge is conserved; the bound state spectrum is of the Coulomb-Kepler type \cite{8}; the scattering cross-section factorizes and has a Rutherford form.

Here we derive similar results in a non-relativistic framework, by enlarging the analogy with the Kepler case and, in particular, with its dynamical symmetries \cite{8,9}.

We first deduce a generalized Pauli equation. For large distances, the $\SU(2)$ symmetry breaks down to $\U(1)$ yielding a conserved electric charge, and we show that the problem reduces to the (at first sight artificial) system
considered by D'Hoker and Vinet \cite{3}, whose Hamiltonian is
\begin{equation}
H=\2(p_i-eA_i)^2-\frac{1}{r}+\frac{q^2}{2r^2}+\frac{1}{2q^2}-
q\sigma_k\frac{x^k}{r^3}.
\label{(1)}
\end{equation}
Here the anomalous gyromagnetic ratio $4$ arises due to (anti)self-duality, and the
potential terms appear as remnants of the long-range Higgs field.
This system admits a large dynamical (super)symmetry, which has been used \cite{3} to derive
the bound state spectrum
\begin{equation}
E=\frac{1}{2q^2}-\frac{1}{2p^2},
\qquad
p=|q|, |q|+1,, \dots
\label{(2)}
\end{equation}
For $p\geq|q|+1$, the multiplicity of the $p$-level is $2(p^2-q^2)$, and for the $E=0$
ground state, it is $2|q|$.

Here we focus our attention on the \emph{scattering}, and calculate the
$S$-matrix in the spin-$1/2$ case by generalizing Zwanziger's method \cite{2}, based on the dynamical
symmetry. We find that the spin-$1/2$  $S$-matrix  \emph{factorizes} into the product of the spinless
$S$-matrix, $S_0$, with a spin-dependent factor.

Zwanziger originally devised his method for the spin-$0$ counterpart of
(\ref{(1)}), whose Hamiltonian is
\begin{equation}
H_0=\2(p_i-eA_i)^2-\frac{1}{r}+\frac{q^2}{2r^2}+\frac{1}{2q^2}.
\label{(3)}
\end{equation}
This system has dynamical symmetries analogous to those in the Kepler problem \cite{8,9}, allowing for an algebraic derivation
of the bound state spectrum
as well as the $S$-matrix, $S_0$ in Eq.~(\ref{(7)}) below,  \cite{7,8,9}.
The dynamical symmetry of (\ref{(3)}) is generated by two conserved vectors, namely by the angular momentum,
\begin{equation}
L^i=\epsilon^{ijk}x^jv^k-q\frac{x^i}{r}
\label{(4)}
\end{equation}
(where $v^k=p^k-eA^k$), and a  ``Runge-Lenz'' vector,
\begin{equation}
\Lambda_0^i=\2\epsilon^{ijk}\big(v^jL^k-L^jv^k\big)-\frac{x^i}{r}.
\label{(5)}
\end{equation}
For fixed value  of  the energy, $E<1/2q^2$, the rescaled Runge-Lenz vector
$\vK=q(1-2q^2 E)^{-1/2}\bLambda_0$ extends the manifest ${\mathrm{o}}(3)$ symmetry into
a dynamical ${\mathrm{o}}(4)$ algebra. The Pauli-Bargmann \cite{8} method yields remarkably the bound state spectrum (\ref{(2)}) once more, but without the ground state $E=0$, and with degeneracy $(p^2-q^2)$, \emph{half} of that of the spinning case.

For energies larger than $1/2q^2$, the vectors $\vJ$ and
$\vK=q(2q^2E-1)^{-1/2}\bLambda_0$
close rather to ${\mathrm{o}}(3,1)$, with Casimirs
\begin{eqnarray}
    \vJ\cdot\vK=\frac{|q|}{k},
\qquad
    \vJ^2-\vK^2=q^2-\frac{1}{k^2}-1,
\label{(6)}
\end{eqnarray}
where $\vk$ is the wave vector and $k=|\vk|$.
Then the algebraic method of Zwanziger \cite{2} yields the $S$-matrix \cite{9},
\begin{equation}
S_0\big(\vl ,\vk\big)=
\big<\vl(\hbox{\small out}),\vk(\hbox{\small in})\big>
=\delta(E_\vl-E_\vk)
\sum_{j\geq |q|}(2j+1) \left[\frac{(j-i/k)!}{(j+i/k)!}\right]
{\cal D}^{\ j\ }_{(-q,q)}\big(R^{-1}_{\vl }R_\vk\big),
\label{(7)}
\end{equation}
where $E_\vk$ is the energy corresponding to
the wave vector $\vk$,
$E_\vk=k^2/2+1/2q^2$ (so that $k=(2E-1/q^2)^{1/2}$), and
$R_\vk$ is a certain rotation matrix which brings the
$z$-axis into
the direction of $\vk$.
Equation (\ref{(7)}) leads to the modified Rutherford cross-section
\begin{equation}
\frac{d\sigma_0}{d\omega}=
\left(\frac{1+(qk)^2}{4k^4}\right){\rm cosec}^4\2\theta\ .
\label{(8)}
\end{equation}

In order to study a spin-$1/2$ particle in a BPS monopole field, let us consider
 a $5$-dimensional extended space-time with metric
$g_{AB}={\rm diag}(g_{\mu\nu},g_{44})={\rm diag}(-1,1,1,1,1)$. Let us chose the representation
$$ 
\gamma^0=\left(\begin{array}{cc}
iI&0\\0&-iI
\end{array}\right),
\qquad
\gamma^k=\left(\begin{array}{cc}
0&i\sigma^k\\-i\sigma^k&0
\end{array}\right),
\qquad
\gamma^4=\left(\begin{array}{cc}
0&I\\I&0
\end{array}\right)
$$
of the Dirac matrices, so that $\left\{\gamma^A,\gamma^B\right\}=2g^{AB}$, $A,B=0,1,2,3,4$.
Consider a Dirac field $\Psi$ which transforms according to an unitary representation
$\Psi\to Q\Psi$ of the gauge group $\SU(2)$, and let $Q_a$ ($a=1,2,3, Q_a=
Q_a^\dagger;\ [Q_a,Q_b]=i\epsilon_{abc}Q_c$)
denote the isospin generators in this representation. To a Yang-Mills-Higgs
configuration $(A_\mu,\Phi)$ with the Higgs field $\Phi$ in the adjoint representation
we can associate a pure YM configuration ${\cal A}_A$ in $5$ dimensions by setting
 ${\cal A}_\mu=A_\mu,\ {\cal A}_4=\Phi$.
 The 5-dimensional Dirac equation
 reads\footnote{Using this equation means that we choose
 the Higgs coupling to be the same as the Yang-Mills coupling.}
\begin{equation}
\Big(\gamma^\mu(\p_\mu+i {\cal A}_\mu^a Q_a)+
\gamma^4(\p_4+i{\cal A}_4^a Q_a) -m\Big)\Psi=0.
\label{(9)}
\end{equation}
Introducing
$$
\vpi = -i \mathbf{\nabla} + Q_a {\mathbf{A}^a},
\qquad
\pi_4=-i\p_4 + Q_a\Phi^a,
\qquad
\Psi=\left(\begin{array}{c}\varphi\\
\chi\end{array}\right),
$$
this equation becomes
\begin{eqnarray}
i\p_t\varphi=\big(\vsigma\cdot\vpi-i\pi_4\big)\chi+A_0^aQ_a \varphi+m\varphi,
\label{(10a)}
\\[8pt]
i\p_t\chi=\big(\vsigma\cdot\vpi+i\pi_4\big)\varphi+A_0^aQ_a \chi
-m\chi.
\label{(10b)}
\end{eqnarray}
In order to take the non-relativistic limit, we introduce, as usual,
 the slowly varying-in-time functions $\psi$ and $\xi$, $\varphi=e^{-imt}\psi$ and
$\chi=e^{-imt}\xi$. These spinors satisfy
\begin{eqnarray*}
i\p_t\psi&=&\big(\vsigma\cdot\vpi-i\pi_4\big)\xi+A_0^aQ_a \psi,
\\[8pt]
i\p_t\xi&=&\big(\vsigma\cdot\vpi+i\pi_4\big)\psi+A_0^aQ_a \xi
-2m\xi.
\end{eqnarray*}
In the non-relativistic limit, the second equation
is approximately solved by $\xi\approx\big(\vsigma\cdot\vpi+i\pi_4\big)\psi/2m$
($\ll \psi$), so that Eq. (\ref{(10a)}) becomes
\begin{eqnarray}
i\p_t\psi=
\left[\smallover1/{2m}\big(\vsigma\cdot\vpi-i\pi_4\big)\big(\vsigma\cdot\vpi+i\pi_4\big)
+A_0^aQ_a\right]\psi.
\label{(11)}
\end{eqnarray}
This equation can be brought into a more familiar form by developing it as
$$
\big(\vsigma\cdot\vpi-i\pi_4\big)\big(\vsigma\cdot\vpi+i\pi_4\big)=
(\vsigma\cdot\vpi)^2+i[\vsigma\cdot\vpi,\pi_4] + \pi_4^2,
$$
and using the identities
\begin{eqnarray*}
(\vsigma\cdot\vpi)^2&=&\vpi^2+\smallover1/4[\sigma_j,\sigma_k][\pi_j,\pi_k]=\vpi^2 +\sigma_kB^a_kQ_a,
\\[8pt]
i[\vsigma\cdot\vpi,\pi_4]&=& \sigma_k(D_k\Phi)=\sigma_k(D_k\Phi)^aQ_a,
\end{eqnarray*}
($B_k=\2\epsilon_{kij}F^{ij}$), which can be verified by a straightforward calculation.
Being extensions from $4$ dimensions, none of the fields depend on the
extra coordinate, so we require $\p_4\psi=0$ too\footnote{A more general possibility would
be to require equivariance, $-i\p_4\psi=\zeta \psi $ with some real constant $\zeta$.}.
 Substitution into Eq. (\ref{(11)}) yields
finally the generalized Pauli equation
\begin{equation}
i\p_t\psi=\smallover1/{2m}\left[\vpi^2+(Q_a\Phi^a)^2+\sigma_k(B_k^a+D_k\Phi^a)Q_a+A_0^aQ_a
\right]\psi.
\label{(12)}
\end{equation}
Similarly, in the spin-$0$ case, one can deduce the Schr\"odinger equation
\begin{equation}
i\p_t\psi=\smallover1/{2m}\left[\vpi^2+(Q_a\Phi^a)^2+A_0^aQ_a
\right]\psi.
\label{(13)}
\end{equation}

So far, we worked with an arbitrary Yang-Mills-Higgs configuration
in Minkowski space. Let us assume henceforth that $(\vA,\Phi)$ is the field of a BPS monopole, i.e., a
static, purely-magnetic ($A_0=0$) solution to the Bogomolny
(i.e., anti-self-duality) equations, $B_k=D_k\Phi$, such that
$\Phi\to1$ as $r\to\infty$. Introducing the
spin matrix $S_k=\sigma_k/2$, the Pauli equation (\ref{(12)}) takes the form
\begin{equation}
i\p_t\psi=\smallover1/{2m}\left[\vpi^2+(Q_a\Phi^a)^2
+4S_kB_k^aQ_a\right]\psi.
\label{(14)}
\end{equation}
Notice that the factor $4$ in Eq. (\ref{(14)}), the anomalous gyromagnetic ratio of the particle, arises due to the Bogomolny equation $B_k=D_k\Phi$.
For the spherically symmetric solution of Bogomolny, and of Prasad-Sommerfield \cite{6}, in particular,
\begin{equation}
\Phi^a = -\displaystyle\frac{x^a}{r}\big(\coth r-\displaystyle\frac{1}{r}\big),
\qquad
A_i^a= \epsilon_{aik} \displaystyle\frac{x^k}{r}\big(1-\displaystyle\frac{r}{\sinh r}\big).
\label{(15)}
\end{equation}

For large distances, the exponentially decreasing terms become small and the fields approach their asymptotic values,
\begin{equation}
\Phi^a=-\displaystyle\frac{x^a}{r}\big(1-\displaystyle\frac{1}{r}\big),
\qquad
A_i^a=\epsilon_{aik} \displaystyle\frac{x^k}{r}.
\label{(16)}
\end{equation}
The Higgs field is in the adjoint representation, so it can be used to define the electromagnetic properties \cite{10}. Indeed, the electric charge operator,
$Q_{em}$, can be introduced as,
\begin{equation}
Q_{em}=\frac{\Phi}{|\Phi|}=
\frac{1}{\sqrt{\Phi^b\Phi^b}}\,\Phi^aQ_a.
\label{(17)}
\end{equation}

All fields are required to be eigenstates of $Q_{em}$, whose eigenvalues
(identified with the electric charge) are integer multiples of a smallest electric charge,
\cite{10}, $q=nq_{min},\ n=0,\pm1,\dots$. If $\Phi/|\Phi|$ is covariantly constant, the
electric charge  operator commutes with the Hamiltonian and the electric charge is conserved. This is what happens asymptotically (i.e., for large $r$) in the BPS case. The gauge potential in Eq. (\ref{(16)}) is in fact a gauge transform of a Dirac monopole
potential $\vA^D$, $(\rot\vA{\strut}^D)^k=x^k/r^3$. (The above BPS solution
has unit magnetic charge $g=1$.)
The electric charge operator is defined hence and is given by
\begin{equation}
Q_{em}=-\frac{x^a}{r}Q_a.
\label{(18)}
\end{equation}
Due to our normalization, $q_{min}\!=\!\2$. Restricted to $Q_{em}\psi=q\psi$ states,
$\Phi^aQ_a\!=\!q(1-1/r),\ B_k^a Q_a=-q x^k/r^3$, and thus $\vpi^2=(-i\p_k-qA_k^D)^2$.
For large distances, the non-Abelian equation (\ref{(14)}) reduces, hence, to
\begin{equation}
i\p_t\psi=\frac{1}{2m}\left[(-i\p_k-qA_k^D)^2-\frac{2q^2}{r}+
\frac{q^2}{r^2}+q^2-2q\sigma_k\frac{x^k}{r^3}\right]\psi.
\label{(19)}
\end{equation}
Rescaling as $\vr\to q^2\vr,\ t\to q^4t$ (and for $m=1$), this is exactly the
D'Hoker-Vinet system (\ref{(1)}).

Similarly, in the spinless case, one gets the Zwanziger system (3).

 Had we studied instead the motion of a spin-1/2 particle in the field of a self-dual
(${\bf D}\Phi=-{\bf B}$) monopole, the Pauli term would \emph{disappear} altogether, and
both spin components would satisfy independent spinless equations yielding
the bound state spectrum (\ref{(2)}) and the $S$-matrix (\ref{(7)}).

In the spin-$1/2$ case, D'Hoker and Vinet \cite{3} show that the system (\ref{(1)}) admits
three conserved vectors, namely
\begin{eqnarray}
J^i&=&L^i+\2\sigma^i
\label{(20a)}
\\[6pt]
\Lambda^i&=&\Lambda_0^i-\epsilon^{ijk}\sigma^jv^k+\frac{q}{r}\sigma^i-
qx^i(\frac{x^j\sigma^j}{r^3})-(\frac{1}{2q})\sigma^i,
\label{(20b)}
\\[6pt]
\Omega^i&=&\2\sigma^iv^2-(\sigma^jv^j)v^i+\epsilon^{ijk}\big(\frac{q}{r}-\frac{1}{q}\big)
\sigma^jv^k-\frac{1}{2}\big(\frac{q}{r}-\frac{1}{q}\big)^2\sigma^i.
\label{(20c)}
\end{eqnarray}
The structure involved becomes clearer if we consider
\begin{eqnarray}
J_1^i&=&J^i+\smallover1/{2H}\Omega^i
\label{(21a)}
\\[6pt]
K^i&=&
(1-2q^2H)^{-1/2}\big(q\Lambda^i-\smallover1/{2H}\Omega^i\big),
\label{(21b)}
\\[6pt]
J_2^i&=&-\smallover1/{2H}\Omega^i.
\label{(21c)}
\end{eqnarray}
For a fixed value $0<E<1/2q^2$ of the energy, the operators
(\ref{(21a)})-(\ref{(21b)})-(\ref{(21c)}) span an
${\mathrm{o}}(4)\oplus{\mathrm{o}}(3)$ dynamical symmetry algebra, which has been used \cite{3} to derive
the bound state spectrum (\ref{(2)}).
For scattering off a monopole, the multiplicities are
actually doubled, because $q$ and $-q$ are just two charge-states of the same particle
with non-Abelian structure. It is worth noting that our non-relativistic
system also exhibits zero-energy bound states found before in the relativistic setup
\cite{5}.

Let us now study those scattering states with fixed energies $E$ larger that $1/2q^2$. $K^i$ is now
\begin{equation}
K^i=
(2q^2E-1)^{-1/2}\big(q\Lambda^i-\smallover1/{2E}\Omega^i\big).
\label{(21b')}
\end{equation}
The commutation relations become
\begin{eqnarray}
    [J_1^j,J_1^k]&=&i\epsilon^{jkl}J_1^l,
    \qquad
    [J_1^j,K^k]=i\epsilon^{jkl}K^l,
    \qquad
    [K^j,K^k]=-i\epsilon^{jkl}J_1^l,
\label{(22)}
\\[10pt]
    [J_2^j,J_2^k]&=&i\epsilon_{jkl}J_2^l,
    \qquad
    [J_2^j,J_1^k]=[J_2^j,K^k]=0,
\label{(22,5)}
\end{eqnarray}
showing that $\vJ_1$ and $\vK$ now span an ${\mathrm{o}}(3,1)$ algebra,
to which $\vJ_2$ adds an independent ${\mathrm{o}}(3)$. Using the formulae in Ref.~\cite{3}, the Casimir
operators are found to be
\begin{equation}
\vJ_1\cdot\vK=\frac{|q|}{k},
\qquad
\vJ_1^2-\vK^2=q^2-\frac{1}{k^2}-1,
\qquad
\vJ_2^2=3/4,
\label{(23)}
\end{equation}
with $k=(2E-1/q^2)^{1/2}$. The Casimirs single out the actual
representation of
${\mathrm{o}}(3,1)\oplus{\mathrm{o}}(3)$ which operates on the scattering states.
The symmetry generators allow for a complete labeling of the states, so we have
an irreducible representation
(spanned, of course, by non-normalizable basis vectors). The ${\mathrm{o}}(3)$ of
$\vJ_2$ is in its spin-$1/2$ representation. Representation theory \cite{11} tells us then that among
the scattering states there is a convenient ``distorted spherical-wave'' basis, $\big|k,j_1,m,\mu\big>$,
defied by the relations

\begin{equation}\begin{array}{lll}
H\big|k,j_1,m,\mu\big>&=&E_k\big|k,j_1,m,\mu\big>,
\qquad
k>0;
\\[8pt]
(\vJ_1)^2\big|k,j_1,m,\mu\big>&=&j_1(j_1+1)
\big|k,j_1,m,\mu\big>,\qquad j_1=|q|,|q|+1,\dots;
\\[8pt]
J_1^3\big|k,j_1,m,\mu\big>&=&m\big|k,j_1,m,\mu\big>,
\qquad
m=-j_1,\dots,j_1;
\\[8pt]
J_2^3\big|k,j_1,m,\mu\big>&=&\mu\big|k,j_1,m,\mu\big> ,
\qquad
\mu=\pm\2;
\\[8pt]
\big<k',j_1',m',\mu'\big|k,j_1,m,\mu\big>&=&
\delta(E_{k'}-E_k)\,\delta_{j_1',j_1}\,\delta_{m',m}\,\delta_{\mu',\mu}\ .
\end{array}
\label{(24)}
\end{equation}
We want to calculate the $S$-matrix
$S(\vl,\sigma|\vk,s)=
\big<\vl ,\sigma (\hbox{\small out})\big|\vk, s(\hbox{\small in})\big>$.
Here $\big|\vk, s(\hbox{\small in})\big>$ and
$\big|\vk, s(\hbox{\small out})\big>$, respectively,
 are solutions of the time-dependent
Pauli equation for the Hamiltonian (\ref{(1)}), which approximate sharply
peaked wave packets incoming and outgoing with velocity
$\vk$ at $t=\mp\infty$. The ``helicity'' of the incoming
(outgoing) particles is fixed to be $s=\pm1/2$, when they are
far away from the scattering center. This means that, for
$t=\mp\infty$, the relations
\begin{equation}\begin{array}{cll}
\vv\,\Big|\vk,s\left(\begin{array}{c}
\hbox{\small in}\\ \hbox{\small out}\end{array}\right)
\Big>
&=&\vk\,
\Big|\vk,s\left(\begin{array}{c}
\hbox{\small in}
\\
\hbox{\small out}
\end{array}\right)\Big> ,
\\[12pt]
\displaystyle\frac{1}{r}\,\Big|\vk,s\left(\begin{array}{c}
\hbox{\small in}\\ \hbox{\small out}\end{array}\right)
\Big>
&=&0 ,
\\[12pt]
\widehat{\vr\,}\,
\Big|\vk,s\left(\begin{array}{c}
\hbox{\small in}\\ \hbox{\small out}\end{array}\right)
\Big>
&=&\mp\ \widehat{\vk}\,
\Big|\vk,s\left(\begin{array}{c}
\hbox{\small in}\\ \hbox{\small out}\end{array}\right)
\Big> ,
\\[12pt]
(\vsigma\cdot\widehat{\vk})
\Big|\vk,s\left(\begin{array}{c}
\hbox{\small in}\\ \hbox{\small out}\end{array}\right)
\Big>
&=&2s
\Big|\vk,s\left(\begin{array}{c}
\hbox{\small in}\\ \hbox{\small out}\end{array}\right)
\Big> ,
\end{array}
\label{(25)}
\end{equation}
with $\widehat{\vr}=\vr/r,\, \widehat{\vk}=\vk/k$,
are verified up to terms which do not contribute to the $S$ matrix.
This implies in turn that our ``distorted plane wave-like'' scattering states can be taken to satisfy
\begin{equation}\begin{array}{cll}
H\,\Big|\vk,s\left(\begin{array}{c}
\hbox{\small in}\\ \hbox{\small out}\end{array}\right)
\Big>
&=&\,
E_k\,\Big|\vk,s\left(\begin{array}{c}
\hbox{\small in}
\\
\hbox{\small out}
\end{array}\right)\Big> ,
\\[12pt]
(\vJ_1\cdot\widehat{\vk})\,\Big|\vk,s\left(\begin{array}{c}
\hbox{\small in}\\ \hbox{\small out}\end{array}\right)
\Big>
&=&
\pm q\Big|\vk,s\left(\begin{array}{c}
\hbox{\small in}\\ \hbox{\small out}\end{array}\right)
\Big> ,
\\[16pt]
(\vK\cdot\widehat{\vk})\,
\Big|\vk,s\left(\begin{array}{c}
\hbox{\small in}\\ \hbox{\small out}\end{array}\right)
\Big>
&=&(i\pm\displaystyle\frac{1}{k})
\Big|\vk,s\left(\begin{array}{c}
\hbox{\small in}\\ \hbox{\small out}\end{array}\right)
\Big>
\\[16pt]
(\vJ_2\cdot\widehat{\vk})\,\Big|\vk,s\left(\begin{array}{c}
\hbox{\small in}\\ \hbox{\small out}\end{array}\right)
\Big>
&=&
s\Big|\vk,s\left(\begin{array}{c}
\hbox{\small in}\\ \hbox{\small out}\end{array}\right)
\Big>
\\[16pt]
\Big<\vl,\sigma\left(\begin{array}{c}
\hbox{\small in}\\ \hbox{\small out}\end{array}\right)
\Big|\vk,s\left(\begin{array}{c}
\hbox{\small in}\\ \hbox{\small out}\end{array}\right)
\Big>
&=&
\delta_{\sigma,s}\,\delta(\vl-\vk).
\end{array}
\label{(26)}
\end{equation}
Alternatively, one is (in principle) able to check these formulae
on explicit scattering wave functions.
Anyway, postulating these crucial relations  allows us to calculate the $S$-matrix.
The method is to first expand
$\big|\vk,s({\rm in})\big>$
(resp. $\big|\vk,s({\rm out})\big>$\ )
in the spherical-wave basis (\ref{(24)})  and to calculate the matrix elements
by using these expansions. This is a practical method since the spherical waves are easy to handle.
Let us choose indeed, for an arbitrary vector $\vk$, a rotation
$R_{\vk}=\exp[i\alpha^a(\vk)T^a]$, which brings the unit-vector
 $\vz$ (directed along the $z$-axis) into the  direction of
$\vk$. Due to spherical symmetry, we can choose scattering states satisfying the phase convention relations
\begin{equation}
\Big|\vk,s\left(\begin{array}{c}
\hbox{\small in}\\ \hbox{\small out}\end{array}\right)
\Big>=\exp [i\alpha^a(\vk)J^a]
\Big|k \vz,s\left(\begin{array}{c}
\hbox{\small in}\\ \hbox{\small out}\end{array}\right)
\Big>\,.
\label{(27)}
\end{equation}
Here the rotation matrices $T^a$ are replaced by the conserved angular momentum components $J^a$, which generate the rotations in the
quantum mechanical state space. As a consequence of the second and fourth Eqs. in  (\ref{(26)}), $\big|k\vz,s({\rm in})\big>$
(resp.  $\big|k\vz,s({\rm out})\big>$ can be expanded as
\begin{equation}
\Big|k\,\vz,s\left(\begin{array}{c}
\hbox{\small in}\\ \hbox{\small out}\end{array}\right)
\Big>
=\sum_{j_1\geq |q|}(2j_1+1)^{1/2}
b_{j_1,s}^{in,out}\,
\Big|k,j_1,m=\pm q,\mu=s\Big>,
\label{(28)}
\end{equation}
and all we have to do is to determine the expansion coefficients.
(The factors $(2j_1+1)^{1/2}$ are included for convenience.)

The third equation in (\ref{(26)}) tells us how
$K^3=\vK\cdot\vz$
acts on the l.h.s. of Eq. (\ref{(28)}).
At the same time, its action on the r.h.s. of (\ref{(28)}) is fixed
by the representation theory. Putting these two pieces of
information together, one gets recursion relations for the coefficients $b$. These
are actually the \emph{same} as in the spin-$0$
case, since the Casimirs of ${\mathrm{o}}(3,1)$ are identical. We can
take, therefore, independently of $s$, the
$b$'s to be those coefficients calculated by Zwanziger \cite{2} for the spinless case,
\begin{equation}
b_{j_1,s}^{in,out}=\left[\frac{(j_1\mp i/k)!}{(j_1\pm i/k)!}
\right]^{1/2}.
\end{equation}
The relations $\vJ=\vJ_1+\vJ_2,\  [\vJ_1,\vJ_2]=0$ imply that the
rotation operator $\exp[i\alpha^a(\vk)J^a]$ factorizes as  $\exp[i\alpha^a(\vk)J^a]=\exp[i\alpha^a(\vk)J_1^a]
\cdot
\exp[i\alpha^a(\vk)J_2^a]$.   This then gives rise to the equation
\begin{equation}
\exp[i\alpha^a(\vk)J^a]\big|k,j_1,m,\mu\big>=
\sum_{m'=-j_1}^{j_1}\sum_{\mu'=\pm 1/2}
\cD_{(m',m)}^{j_1}(R_{\vk})
\cD_{(\mu',\mu)}^{1/2}(R_{\vk})\big|k,j_1,m',\mu'\big>,
\label{(30)}
\end{equation}
with the rotation matrices $\cD$ defined by
$\exp[i\alpha^a(\vk)J_1^a]$ and
$\exp[i\alpha^a(\vk)J_2^a]$ in the representations belonging to spin-$j_1$ and spin-$1/2$, respectively.
Collecting our results, we obtain the $S$-matrix
\begin{equation}
S\big(\vl,\sigma,\vk,s\big)=
\delta\big(E_{\vl}-E_{\vk}\big)\cD^{1/2}_{(s,\sigma)}
\big(R_{\vl}^{-1}R_{\vk}\big)
\\[12pt]
\sum_{j\geq |q|}(2j+1)\left[\frac{(j-i/k)!}{(j+i/k)!}\right]
{\cal D}^{\ j\ }_{(-q,q)}\big(R^{-1}_{\vl}R_\vk\big),
\label{(31)}
\end{equation}
showing that the spin and momentum dependence indeed factorizes.
This equation can be rewritten in fact as,
\begin{equation}
S(\vl,\sigma|\vk,s)=\cD^{1/2}_{(s,\sigma)}
\big(R_{\vl}^{-1}R_{\vk}\big)\cdot
S_0(\vl|\vk).
\label{(32)}
\end{equation}
We see that $S$ and $S_0$ have the same poles, namely at $j-i/k=-n,\ n=1,2,\dots$, yielding the positive-energy
bound state spectrum (\ref{(2)})
with $n=n+j=|q|+1,\dots$. The zero-energy bound states do not appear
in the $S$ matrix. Equation (\ref{(32)}) implies that, for a particle with incoming helicity $s$ and outgoing helicity $s'$, the scattering cross-section is,
\begin{equation}
\left(\frac{d\sigma}{d\omega}\right)_{s',s}=
\frac{d\sigma_0}{d\omega}\,\big|{\bf d}^{1/2}_{s',s}(\theta)|^2,
\label{(33)}
\end{equation}
where ${\bf d}^{1/2}_{s',s}(\theta)$ is the matrix of a rotation by angle $\theta$ around the $y$-axis in
the representation spin-$1/2$.
Since $s$, $s'=\pm1/2$, ${\bf d}^{1/2}_{1/2,1/2}(\theta)=
{\bf d}^{1/2}_{-1/2,-1/2}(\theta)=\cos\2\theta$,
${\bf d}^{1/2}_{-1/2,1/2}(\theta)=
{\bf d}^{1/2}_{1/2,-1/2}(\theta)=\sin\2\theta$, the total cross section
\begin{equation}
\left(\frac{d\sigma}{d\omega}\right)_{\hbox{\small total}}=
\frac{1}{2}
\sum
\left(\frac{d\sigma}{d\omega}\right)_{s',s}
=\frac{d\sigma_0}{d\omega},
\label{(34)}
\end{equation}
is \emph{identical} to that in Eq. (\ref{(8)}) for spin-$0$. Remarkably, the results are
essentially the same as in the relativistic treatment \cite{5}. The factorization
of the $S$-matrix is actually a general property, explained by supersymmetry \cite{12}.

\kikezd{ACKNOWLEDGEMENTS}.

We are indebted to Z. Horvath and L. O'Raifeartaigh  for discussions.

\kikezd{Note added.}
Neither the original text nor the list of references,
published as {\sl Mod. Phys. Lett.} {\bf A3} No 15.  1451-1460 (1988), have been updated.

\end{document}